\documentclass[prl,twocolumn]{revtex4}

\usepackage{amssymb}
\usepackage{amsmath}
\usepackage{paralist}
\usepackage{graphicx}
\usepackage{xcolor}
\usepackage{fancyhdr}
\usepackage{xr}

\newcommand{\ii}{{\rm i}}
\newcommand{\e}{{\rm e}}

\begin{document}
%
%
    \title{Perfect Transmission through Disordered Media}
%
%
    \author{C. G. King}
    \author{S. A. R. Horsley}
    \author{T. G. Philbin}
    \affiliation{Department of Physics and Astronomy, University of Exeter, Stocker Road, Exeter, EX4 4QL}
%
%
    \begin{abstract}
    The transmission of a wave through a randomly chosen `pile of plates' typically decreases exponentially with the number of plates, a phenomenon closely related to Anderson localisation. In apparent contradiction we construct disordered planar permittivity profiles which are complex-valued (i.e. have reactive and dissipative properties) that appear to vary randomly with position, yet are one-way reflectionless for all angles of incidence and exhibit a transmission coefficient of unity.  In addition to these complex-valued 'random' planar permittivity profiles, we construct a family of real-valued, two-way reflectionless and perfectly transmitting disordered permittivity profiles that function only for a single angle of incidence and a narrow frequency range.
    \end{abstract}
    \maketitle
%
%
    A wave propagating through \(N\) randomly chosen lossless slabs of material tends to be exponentially extinguished as \(N\) increases~\cite{Berry1997,Baluni1985}. 
The transmission through such a random combination of slabs is given by the geometric mean of the transmissivity, \(|t_{\text{eff}}|^{2}=\exp(2\left<\text{log}(|t|)\right>)\) corresponding to averaging over all possible realisations. For \(N\) slabs this is~\cite{Berry1997}
    \begin{equation}
        |t_{\text{eff}}|^{2}=\text{exp}\left(-2\sum_{i=1}^{N}\left<\text{log}\left(\frac{1}{|t_{i}|}\right)\right>\right),\label{1}
    \end{equation}
where $t_{i}$ is the transmission coefficient for the $i^{\rm th}$ slab)~\cite{footnote1}. The average transmissivity (\ref{1}) clearly decreases exponentially with increasing \(N\) (see~\cite{Lu2009} for bounds), leading to the phenomenon where a layered transparent disordered medium tends to act as a good mirror. In this Letter, we construct families of layered media with permittivities that are similarly random in the direction of propagation and yet the expected high reflection and low transmission is avoided. The layered media we explore all exhibit disorder and are described by a permittivity that is a continuous function of position $\epsilon(x)$.
    \par
    The overwhelming majority of possible profiles are disordered, and the exponentially small value of the geometric mean of the random phase model (\ref{1}) can be seen as stemming from the dominant contribution of disordered media to the average.  It is also connected to the phenomenon of Anderson (strong) localisation, which predicts that the eigenstates of a given disordered lattice will tend not to extend over the entire lattice, but will be localised around each of the sites~\cite{Anderson1958,Anderson1977}.  Our perfectly transmitting disordered profiles are in contrast with the many ordered media with unit transmission such as the P{\"o}schl-Teller potential~\cite{Lekner2007}.  We note that while nearly every finite profile is perfectly transmitting for some particular frequency, for disordered media this tends to be over a very narrow band of frequencies.  Here we demonstrate how to design disordered media that transmit close to all the incident light over a comparatively large bandwidth.  
    \par
    The degree of order of our profiles is quantified through examining the average behaviour of the two-point correlation function \(g(s)\) of $\epsilon(x)$.  Ordered media are such that \(g(s)\) is significantly different from zero far away from \(s=0\).  Given information about only the correlation function and Hurst exponent for the profiles discussed in this work, one would conclude that the material parameters were generated from a random-walk like process. Indeed we show that permittivity profiles can admit arbitrarily small correlations (i.e. arbitrarily high levels of disorder), yet they exhibit unit transmission and zero reflection by design.  This is inspired by the recent findings of Yu and co-workers~\cite{Yu2015}, who showed that random-walk like permittivity profiles \(\epsilon(x)\) can be generated through applying a sequence of supersymmetric transformations to an ordered profile, with the disordered profile inheriting the reflectivity and transmissivity from the ordered one.  Our disordered structures are not derived from ordered ones, but in one case are quite an extreme example of profiles that satisfy the spatial Kramers-Kronig relations~\cite{Horsley2015,Longhi2015,Philbin2015,Longhi2015(2),Horsley2016}, a recently highlighted relationship between the real and imaginary parts of \(\epsilon(x)\) that can be used to guarantee zero reflection as well as unit transmission. In another case we construct a real valued disordered profile \(\epsilon(x)\) that is reflectionless and perfectly transmitting, using an ansatz previously used by Berry and Howls~\cite{Berry1990}. While it is unlikely that these combinations of material parameters occur in nature, it may be possible to explore these structures with metamaterials where such a precise control of the disorder may be useful for the construction of mirrors which reflect well over a wide range of frequencies, except over some particular designed band.
    \par
    Consider the 2-dimensional situation shown in figure~\ref{model} for propagation of electromagnetic waves through a slab of material sitting in free space, inhomogeneous along one spatial direction.
    \begin{figure}[ht!]
        \begin{center}
	\includegraphics[width=\linewidth]{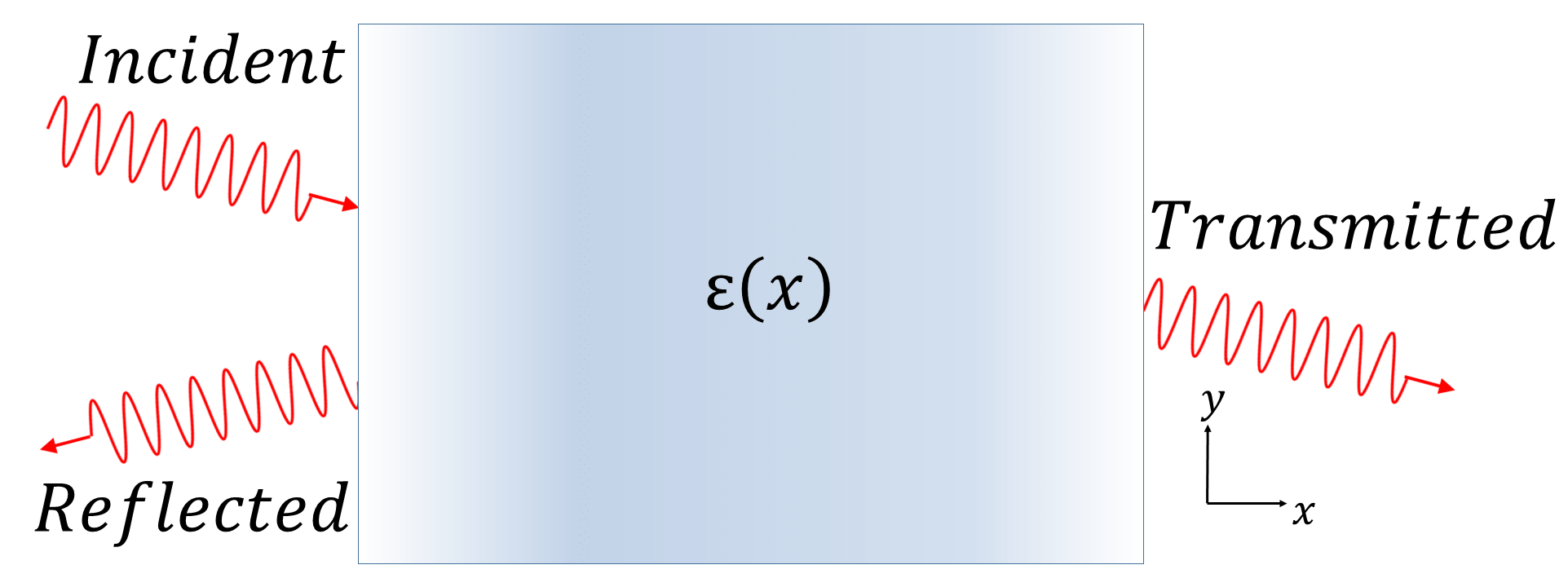}
        \caption{A wave of wavevector \(\textbf{k}=(k_{x},k_{y},0)\) is incident on a material inhomogeneous along the \(x\)-axis with permittivity \(\epsilon(x)\), where \(\epsilon(x)\to1\) as \(|x|\to\infty\). The reflection coefficient is determined by the variation of \(\epsilon\) in space.\label{model}}
        \end{center}
    \end{figure}
As a consequence of Maxwell's equations, the propagation of TE-polarised electromagnetic radiation through dielectric media is governed by the following linear wave equation for the out of plane component of the electric field, \(\varphi\):
    \begin{equation}
        \left[\frac{d^{2}}{dx^{2}}+k_{0}^{2}\epsilon(x)-k_{y}^{2}\right]\varphi(x)=0.\label{2}
    \end{equation}
We have assumed that the magnetic permeability of the slab is unity, and throughout the remainder of this work we take \(k_{y}=0\).
    \par
For a general spatial variation of \(\epsilon(x)\), there is no useful exact solution to (\ref{2}). However, it is possible to find some quite general results concerning wave reflection and transmission through considering the behaviour of the permittivity \(\epsilon(z)\) as a function of complex position \(z=x_{1}+\ii x_{2}\) (see e.g.~\cite{Horsley2015,Horsley2016}). Specifically, writing \(\epsilon(z)=\epsilon_{b}+\chi(z)\), it is known that if \(\chi(z)\) satisfies the Kramers-Kronig relations in space
    \begin{equation}
    \begin{split}
        \text{Re}(\chi(x))&=\frac{1}{\pi}\mathbb{P}\int_{-\infty}^{\infty}\frac{\text{Im}(\chi(x'))}{x'-x}dx'\\
        \text{Im}(\chi(x))&=-\frac{1}{\pi}\mathbb{P}\int_{-\infty}^{\infty}\frac{\text{Re}(\chi(x'))}{x'-x}dx'\label{3},
    \end{split}
    \end{equation}
then there is no reflection for waves incident from the left at any angle of incidence, a result which has recently been verified experimentally using metamaterials~\cite{Jiang2017}. A proof of this result may be found in~\cite{Horsley2015} whereby the scattered field is written in terms of the Fourier transform of a potential. For Kramers-Kronig media, this potential has only positive Fourier components, thus prohibiting the conversion of a positive \(k_{x}\) right-propagating wave into a negative \(k_{x}\) left-propagating wave. However, the result is of sufficient importance to this work that we give an alternative explanation of the result in the Supplementary Material~\cite{Supplementary}. In particular, the susceptibility, \(\chi\), satisfying the spatial Kramers-Kronig relations (\ref{3}) is equivalent to \(\chi\) being an analytic function of position, \(z\), in the upper half \(z\) plane. In addition to the reflectionlessness of Kramers-Kronig media, their transmission coefficient is given by~\cite{Longhi2015,Horsley2016}
    \begin{equation}
        |t|^{2}=\e^{-2k_{0}\text{Im}\left(\int_{-\infty}^{\infty}\sqrt{\epsilon(x)}dx\right)}.\label{4}
    \end{equation}
The result (\ref{4}) implies that if the permittivity is formed as a sum of poles in the lower half complex position plane~\cite{Horsley2016}
    \begin{equation}
        \chi(z)=\sum_{k=1}^{N_{1}}\frac{a_{1,k}}{z-z_{1,k}}+\sum_{k=1}^{N_{2}}\frac{a_{2,k}}{(z-z_{2,k})^{2}}+...\label{5}
    \end{equation}
(thus automatically satisfying (\ref{3})), the transmission coefficient is equal to
    \begin{equation}
        |t|=\e^{\frac{1}{2}\pi k_{0}\text{Re}\sum_{k=1}^{n}a_{1,k}}\label{6}
    \end{equation}
and depends only on the residues of the simple poles in (\ref{5}). Importantly this implies that such profiles consisting of poles of order two or higher (so \(N_{1}=0\) in (\ref{5})) exhibit zero reflection and perfect transmission \textit{regardless} of their number \(N_{j}\), weight \(a_{j,k}\) or position \(z_{j,k}\) in the lower half plane (\(\text{Im}(z_{j,k})<0\)). See~\cite{Horsley2015} for a simulation of the wave propagation through the permittivity corresponding to a single double pole on the negative imaginary axis. More generally, perfect transmission can be achieved when the complex function \(\chi(x)\) both satisfies (\ref{3}) and integrates to zero over the real line,
    \begin{equation}
        \int_{-\infty}^{\infty}\chi(x)dx=0\label{7}
    \end{equation}
a requirement found in~\cite{Longhi2015,Horsley2016} and referred to by Longhi as the 'cancellation condition' (see~\cite{Kober1942}). For profiles given by (\ref{5}), the cancellation condition is equivalent to only having poles of order two, or higher, and is guaranteed to give perfect transmission.
    \par
    It is important to stress just how large this family of media are. In particular, there is enough freedom to allow for the construction of complex valued disordered media exhibiting perfect transmission. As we shall show, the transmission can remain unity while the real and imaginary parts of the permittivity can both be arranged to possess two-point correlation functions
    \begin{equation}
        g(s)=\frac{\int_{-\infty}^{\infty}\chi(x)\chi^{*}(x+s)dx}{\int_{-\infty}^{\infty}\left|\chi(x)\right|^{2}dx}\label{8}
    \end{equation}
that are small except for \(s\sim 0\) and with a Hurst exponent~\cite{Yu2015,Hurst1951} close to 0.5 (indicating a random walk-like character, see supplementary material~\cite{Supplementary}). 

\textit{Perfect transmission through real-valued profiles:}
    Before treating the disordered complex media discussed above, for comparison we construct a family of real disordered profiles that have a rapidly decaying two-point correlation function and a Hurst exponent close to 0.5, but yet exhibit unit transmission. To do this we apply the technique of Berry and Howls~\cite{Berry1990}, where the following ansatz for \(\varphi\) is substituted into equation (\ref{2})
    \begin{equation}
        \varphi(x)=\frac{1}{p(x)^{1/4}}\text{exp}\left(\pm\ii \kappa\int^{x}dx'\sqrt{p(x')}\right)\label{9}
    \end{equation}
which is based on the form of the WKB solutions given in e.g.~\cite{Heading}. The two possible signs in the exponent correspond to right and left travelling waves propagating without reflection, with unit transmission when \(p(x)\) is real and tending to 1 at \(x\to\pm\infty\). Upon substitution of (\ref{9}) into the Helmholtz equation (\ref{2}), one can solve for the requisite permittivity profile, which is found to be
    \begin{equation}
        \epsilon(x,\kappa)=p(x)-\frac{p(x)^{1/4}}{\kappa^{2}}\frac{d^{2}}{dx^{2}}\left(\frac{1}{p(x)^{1/4}}\right).\label{10}
    \end{equation}
Equation (\ref{10}) gives a recipe for the construction of real-valued permittivity profiles that are reflectionless at normal incidence, for fixed \(\kappa\). By choosing \(p(x)\) as a randomly varying function (with a rapidly decaying two-point correlation function and a Hurst exponent close to 0.5), we obtain a similarly randomly varying permittivity profile that exhibits perfect transmission at the wavenumber \(k_{0}=\kappa\). An example of this is shown in figure~\ref{BerryHowl}, where \(p(x)\) is defined as a finite (but long) Fourier sine series with compact support
    \begin{equation}
        p(x)=1+
    \begin{cases}
        \sum_{n=1}^{N}a_{n}\text{sin}\left(\frac{n\pi x}{L}\right),\qquad 0<x<L\\
        0,\qquad\qquad\qquad\qquad\qquad\text{otherwise}\label{11}
    \end{cases}
    \end{equation}
and the coefficients are chosen randomly in such a way that \(\sum_{i=1}^{\lfloor{\frac{n-1}{2}\rfloor}}(n-2i)a_{n-2i}\) is taken from a symmetric real uniform distribution for \(n=1,2,...,N-2\) and is vanishing for \(n=N-1,N\). This ensures the smoothness of \(p\), and hence the continuity of \(\epsilon\).
    \begin{figure}[ht!]
        \begin{center}
	\includegraphics[width=\linewidth]{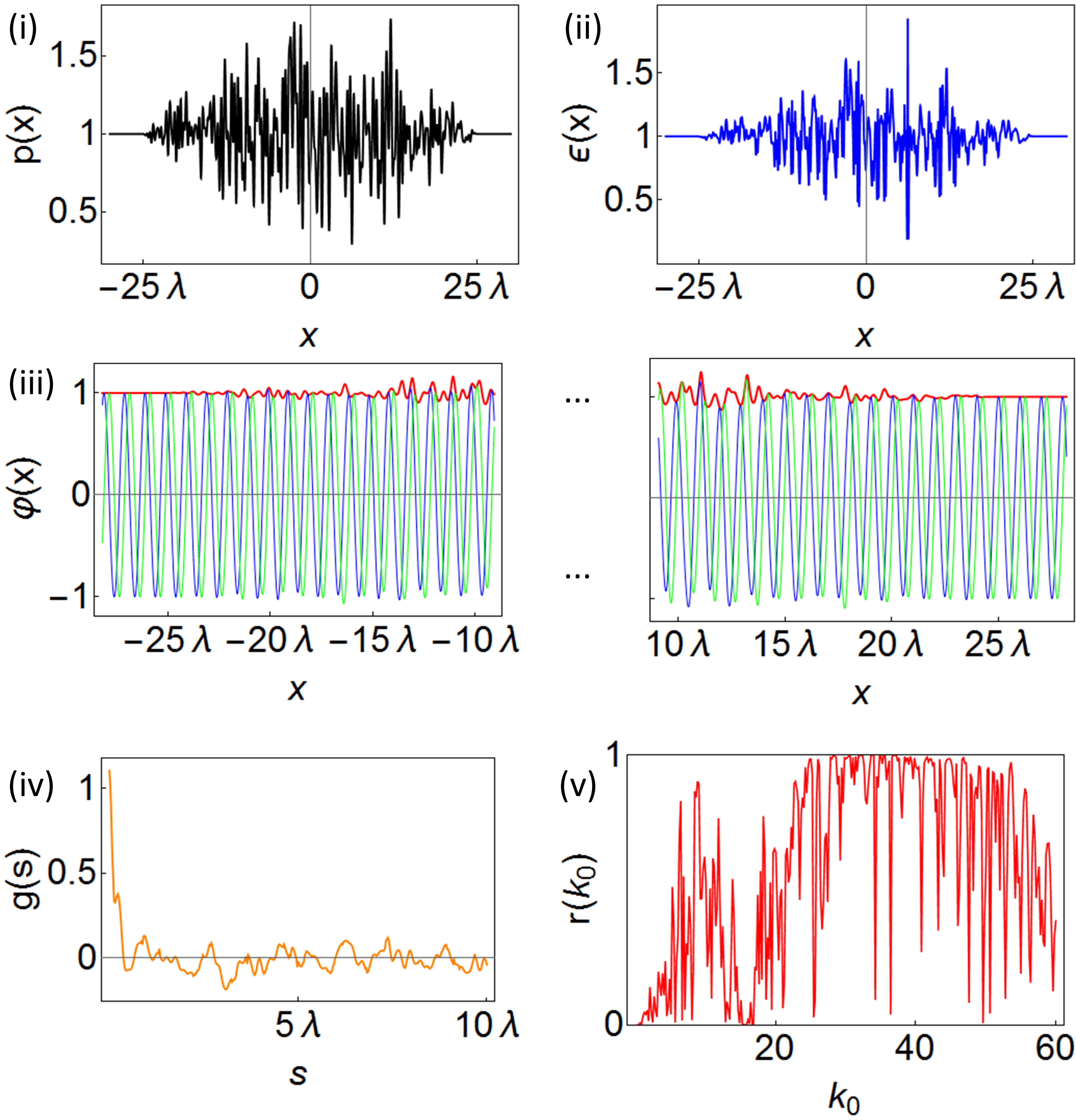}
        \caption{(i) A particular choice of \(p(x)\) given by (\ref{11}) with \(N=250\). (ii) The corresponding permittivity when \(\kappa=5\pi\), having Hurst exponent \(H=0.501\). (iii) Time-averaged (red, upper) amplitude of a left incident wave of wavenumber \(\kappa\), propagating through a medium with permittivity \(\epsilon(x,\kappa)\). Real and imaginary parts of the wave are shown in blue and green, respectively. The lack of oscillations in (iii) indicates that the profile is reflectionless for this wave. The wave is also transmitted without a change in amplitude or a shift in phase. (iv) The profile's correlation function, with correlation length \(x_{L}=0.920\lambda\). (v) The reflection as a function of wavenumber \(k_{0}\). The reflection coefficient is very sensitive to the frequency of the incident wave and high reflection is possible except in a region about \(k_{0}=\kappa\) where the reflection is negligible.\label{BerryHowl}}
        \end{center}
    \end{figure}
    \par
Creating a structure like that shown in figure~\ref{BerryHowl}(ii) in the lab to a high degree of accuracy presents a significant practical challenge and inevitably there will always be some error in the material's permittivity, and, the larger the error, the greater the reflection from the material~\cite{Supplementary}.

   \textit{Perfect transmission through complex-valued profiles:}
    Given a continuous real-valued function, \(\chi_{R}(x)\), satisfying \(\int_{-\infty}^{\infty}\chi_{R}(x)dx=0\), with compact support (corresponding to a finite length medium sitting in vacuum), representing the real part of the susceptibility, the corresponding imaginary part \(\chi_{I}(x)\) belongs to \(L^{1}(-\infty,\infty)\) with vanishing integral on the real line, a consequence of the spatial Kramers-Kronig relations (\ref{3}). Hence, the full susceptibility satisfies the cancellation condition~\cite{Kober1942}. In particular, \(\chi(x)<O\left(\frac{1}{x}\right)\) as \(x\to\pm\infty\) and as a result the corresponding permittivity satisfies \(\text{Im}\left(\int_{-\infty}^{\infty}\sqrt{\epsilon(x)}dx\right)=0\) allowing for perfect transmission.
    \par
Continuing from the previous example, consider the Fourier sine series for the susceptibility:
    \begin{equation}
        \chi_{R}(x)=
    \begin{cases}
        \sum_{n=1}^{N}a_{n}\text{sin}\left(\frac{n\pi x}{L}\right),\qquad 0<x<L\\
        0,\qquad\qquad\qquad\qquad\qquad\text{otherwise}\label{12}
    \end{cases}
    \end{equation}
but now the coefficients, \(a_{n}\), are chosen randomly from a uniform distribution centred around \(0\) for \(n>2\) and \(a_{1}=-\sum_{m=1}^{\lfloor{\frac{N-1}{2}\rfloor}}\frac{a_{2m+1}}{2m+1}\) to ensure that \(\int_{-\infty}^{\infty}\chi_{R}(x)dx=0\). An example of permittivity profiles constructed from a Fourier sine series are shown in figure~\ref{Fourier}.
    \begin{figure}[ht!]
        \centering
	\includegraphics[width=\linewidth]{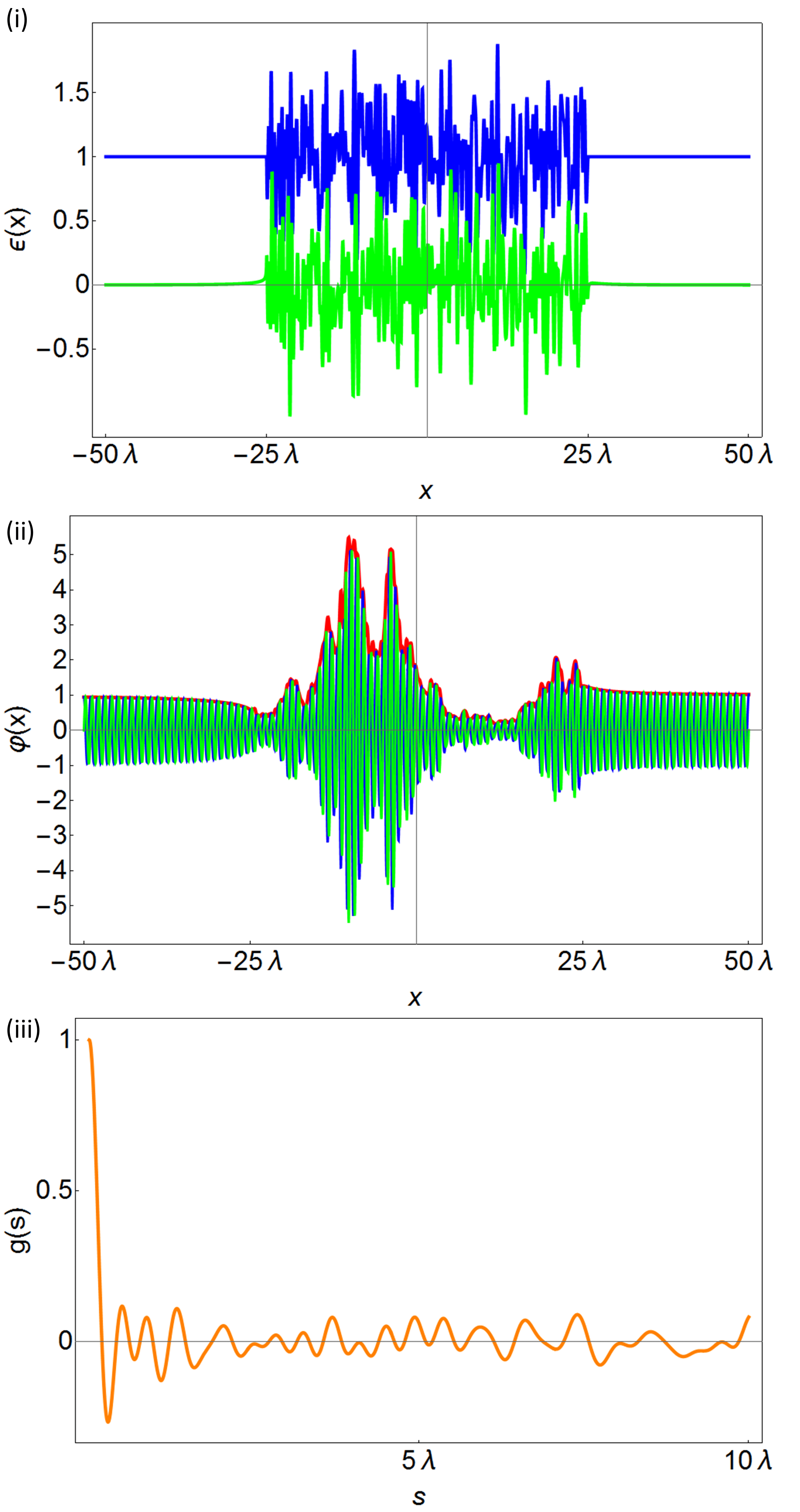}
        \caption{(i): A permittivity profile given by (\ref{12}) with \(N=250\). (ii) The corresponding time-averaged amplitude (red) for a  left incident wave, \(\varphi\), with the real and imaginary parts of the wave are shown in blue and green, respectively. Again, the solution is seen to be reflectionless with perfect transmission. (iii) The correlation function of the permittivity, with correlation length \(x_{L}=0.136\lambda\). Its Hurst exponents are \(H=0.587\) (real part) and \(0.651\) (imaginary part).\label{Fourier}}
    \end{figure}
Unlike the previous example, these spatial Kramers-Kronig profiles transmit perfectly for all angles of incidence (graphs to show this may be found in the Supplementary Material~\cite{Supplementary}).
    \par
Although individual profiles can always have some correlation regardless of the number of terms, \(N\), the average, \(\mu\), and variance \(\sigma^{2}\) of correlations over different realisations generated in this way can be calculated to be diminishing as \(N\to\infty\) (see Supplementary Material~\cite{Supplementary}), thus showing that any correlations will disappear as more terms in the series are taken. Also, there is no obvious way of determining the method of generating the permittivity from a plot of the correlation function. Other methods, such as building the profile as a series of double poles in the lower half position plane lead to a similarly noisy correlation function. However, there may be differences in other measures, such as the Hurst exponent (again see the supplementary material~\cite{Supplementary} for a fuller discussion). The reason why localisation has been avoided relates to the intricate connection between the real and imaginary parts of the permittivity (\ref{3}). By means of a direct substitution of the Hilbert transform (\ref{3}) and a careful rearranging of the principal-value integrals, it can be shown that
    \begin{equation}
        \int_{-\infty}^{\infty}\chi_{I}(y)\chi_{I}(y+s)dy=\int_{-\infty}^{\infty}\chi_{R}(y)\chi_{R}(y+s)dy\label{13}
    \end{equation}
and hence that the correlation function of the real and imaginary parts of the susceptibility are equivalent and also coincide with the real part of the correlation function of the full complex susceptibility. Therefore, given a function to represent the real part of a susceptibility, another function with the same correlation can be found to represent the imaginary part to give a reflectionless perfectly transmitting susceptibility.

\textit{Summary and Conclusions:}
    In this work we have explored a family of disordered permittivity profiles that appear to have been generated by a random walk-like process yet exhibit unit transmission and zero reflection, thus avoiding Anderson localisation. We constructed a family of such real valued profiles, finding that the unit transmission is restricted to a narrow band of frequency and a particular incidence angle. In contrast we examined another family of complex valued profiles where the real and imaginary parts independently appear to have been generated by a random walk-like process, but are related to each other by the spatial Kramers-Kronig relations. Although the construction of such profiles represents an enormous practical challenge, we found that they exhibit perfect transmission whatever the frequency or angle of incidence, remaining perfectly transparent. The level of disorder in the profiles has been measured using the correlation length and Hurst exponent and can take a wide range of values, indicating a wide range of randomness is possible.

\textit{Acknowledgements:}
CGK acknowledges financial support from the EPSRC Centre for Doctoral Training in Electromagnetic Metamaterials EP/L015331/1. SARH acknowledges financial support from  EPSRC program grant EP/I034548/1, the Royal Society and TATA. TGP acknowledges financial support from EPSRC program grant EP/I034548/1. The authors acknowledge useful discussions about localisation in disordered media with J Bertolotti.

\renewcommand{\thepage}{S\arabic{page}}  
\renewcommand{\thesection}{S\arabic{section}}   
\renewcommand{\thetable}{S\arabic{table}}   
\renewcommand{\thefigure}{S\arabic{figure}}
\renewcommand*{\bibnumfmt}[1]{[S#1]}

    \clearpage
    In this supplementary material we briefly explain why spatial Kramers-Kronig profiles are one-way reflectionless, discuss two measures of randomness for functions, describe the sensitivity of the permittivity profiles reflectionless property to deviations from the exact design and consider the capability of these perfectly transmitting media to perform at different frequencies and incident angles. This is followed by an explanation of why the mean and variance of the correlation function of a permittivity profile derived from a Fourier series tends to zero as the number of terms in the series tends to infinity. Finally we discuss the differences between two families of disordered Kramers-Kronig permittivity profiles both being reflectionless with unit transmission; one constructed from poles in the lower half position plane and the other constructed from a Fourier series real part together with its corresponding Hilbert transform imaginary part.
    \maketitle
%
%

\textit{Spatial Kramers-Kronig relations and the reflectionlessness of left incident waves:}
Consider the permittivity appearing in Helmholtz's equation (\ref{2}) being analytically continued to a complex position \(z=x_{1}+\ii x_{2}\), \(\epsilon(z)=1+\chi(z)\). There are a special class of profiles that are analytic on one half of the complex position plane~\cite{Horsley2015,Horsley2016}. These satisfy the spatial Kramers-Kronig relations
    \begin{equation}
    \begin{split}
        \text{Re}(\chi(x))&=\frac{1}{\pi}\mathbb{P}\int_{-\infty}^{\infty}\frac{\text{Im}(\chi(x'))}{x'-x}dx'\\
        \text{Im}(\chi(x))&=-\frac{1}{\pi}\mathbb{P}\int_{-\infty}^{\infty}\frac{\text{Re}(\chi(x'))}{x'-x}dx'\label{S1}.
    \end{split}
    \end{equation}
The derivation of this result may be found in~\cite{Landau}.
    \par
\begin{figure}[ht!]
        \begin{center}
	\includegraphics[width=\linewidth]{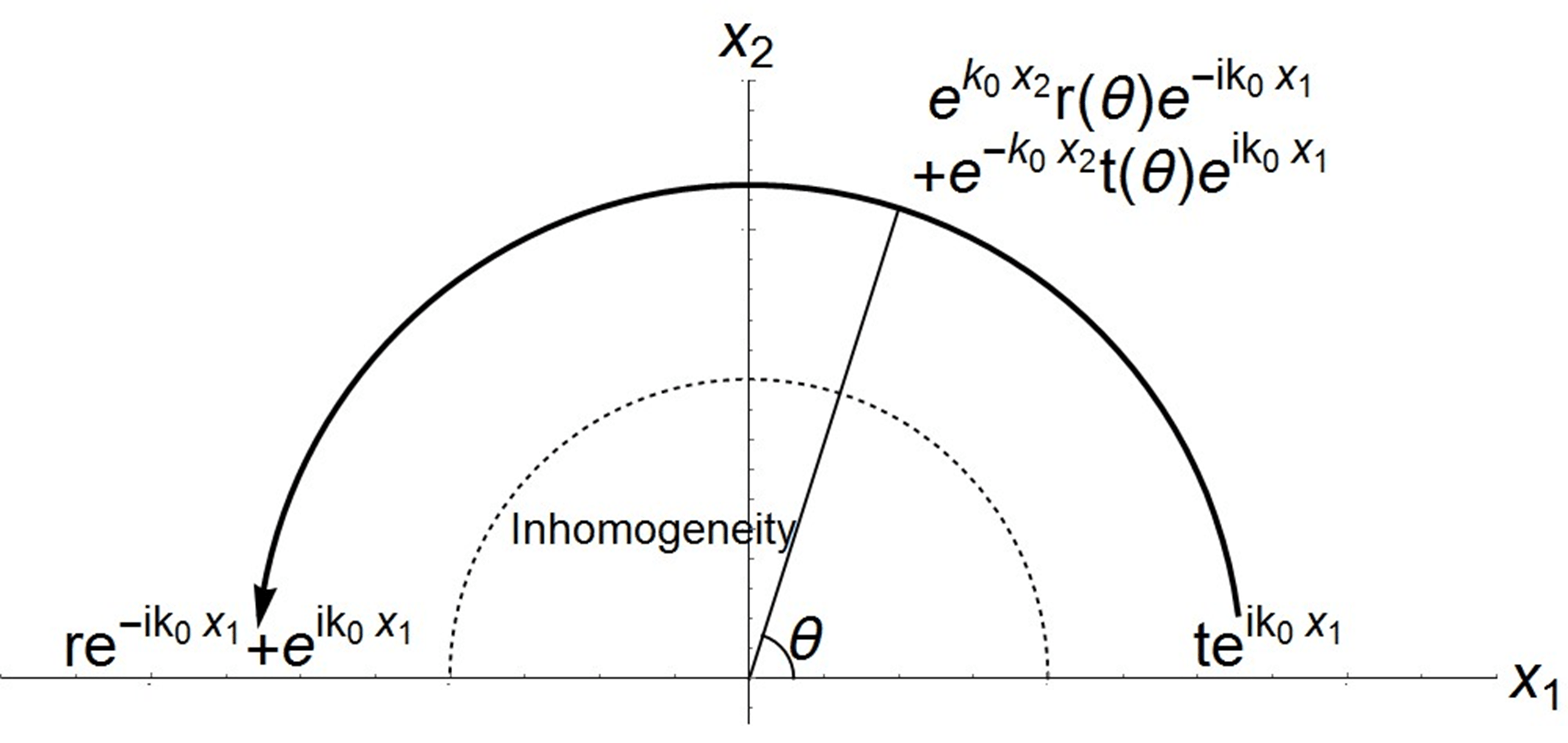}
        \caption{On the far right of the profile, the asymptotic form of the wave can be analytically continued into the upper half position plane. The right-going wave is exponentially diminished while the left-going wave is exponentially amplified.\label{WKB}}
        \end{center}
    \end{figure}

Consider a wave propagating left-to-right through the medium. The analytic continuation of the transmitted wave along the large semi-circle of the complex position plane is shown in figure~\ref{WKB}.
The asymptotic behaviour of the solution in vacuum at an angle \(\theta\) can be written as a combination of left and right propagating waves \(t(\theta)\e^{\ii k_{0}z}+r(\theta)\e^{-\ii k_{0}z}\). A non-zero reflection coefficient \(r(\theta)\) leads to an exponentially growing solution as the semi-circle radius is increased. However, as the solution must be analytic in this upper half plane, and the susceptibility decays to zero, there cannot be a discontinuity in the asymptotic behaviour of the solution, which would be required by a non-zero reflection coefficient, thus explaining why the reflection coefficient must vanish.
    \par
Without the requirement of analyticity in the upper half plane, this argument breaks down due to the presence of branch cuts crossing the semicircular path of figure~\ref{WKB}, across which the asymptotic expansion of the solution in terms of plane waves is discontinuous~\cite{Horsley2015}. This leads to a Stokes phenomenon- the presence of differing asymptotic expansions in different regions of the complex plane~\cite{Heading}. Having this analyticity condition removes the Stokes phenomenon, and hence any reflected wave.
    \par
By replacing the left and right propagating plane waves with the more accurate WKB waves
    \begin{equation}
        \e^{\pm\ii k_{0}\int_{a}^{z}\sqrt{\epsilon(\hat{z})}d\hat{z}}\label{S2}
    \end{equation}
and keeping track of the zero phase reference point, \(a\), as it moves along the semi-circle with the solution, the transmission coefficient can be calculated, in the limit as semi-circle radius tends to infinity, as
    \begin{equation}
        t=\e^{\ii k_{0}\int_{-\infty}^{\infty}\sqrt{\epsilon(x)}dx}\label{S3}.
    \end{equation}
Therefore, a transmission coefficient of unity amplitude (perfect transmission) is obtained when Im\(\left(\int_{-\infty}^{\infty}\sqrt{\epsilon(x)}dx\right)=0\), or equivalently when
    \begin{equation}
        \text{Im}\left(\int_{-\infty}^{\infty}\chi(x)dx\right)=0\label{S4}.
    \end{equation}

\textit{Measures of disorder: The averaged two-point correlation function and the correlation length:}
    The correlation length describes the (in this case, spatial) extent to which the value of a function at one point determines its value at other points. Given a complex-valued function \(f\) belonging to \(L^{2}[a,b]\), its correlation function is defined as
    \begin{equation}
        g_{f}(s)=\frac{\int_{a}^{b-s}f(x)f^{*}(x+s)dx}{\int_{a}^{b}\left|f(x)\right|^{2}dx}\label{S5}
    \end{equation}
where \(s>0\) ([a,b] will be all of the real line for the purpose of this work). The denominator acts as a normalisation factor, ensuring that the correlation function begins at \(g_{f}(0)=1\) and does not exceed unity in absolute value. For a typical real-valued fluctuating function, \(g_{f}(s)\) decays to zero over some length scale, which we shall call the correlation length. This can be naturally defined as the smallest value of \(s\), \(x_{L}\), satisfying \(g_{f}(s)=\text{exp}(-1)\).
    \par
For complex-valued functions, the correlation functions and correlation lengths can be calculated for the real and imaginary parts of \(f\) separately. To include the cross-correlation between the real and imaginary parts, one should, however, take the correlation function of the full function and can then obtain a correlation length from its real part. In general, these may differ significantly, although for a function satisfying the Kramers-Kronig relations, the correlation functions for the real and imaginary parts of \(f\) coincide with the real part of the correlation function of \(f\).

\textit{Measures of disorder: The Hurst exponent:}
   This is an additional tool which can be used to classify the type of randomness exhibited by a disordered function, which is not be identifiable from the correlation function. The details of the calculation are outlined in e.g.~\cite{Yu2015}. The Hurst exponent is a property of sequences, so it is necessary to translate the function describing the permittivity into a sequence using an appropriate discretisation. The 'discretisation width' is chosen to be small compared to the correlation length. A value of H greater (smaller) than 0.5 indicates long term positive (negative) correlations in the sequence.

\begin{figure}[ht!]
        \centering
	\includegraphics[width=\linewidth]{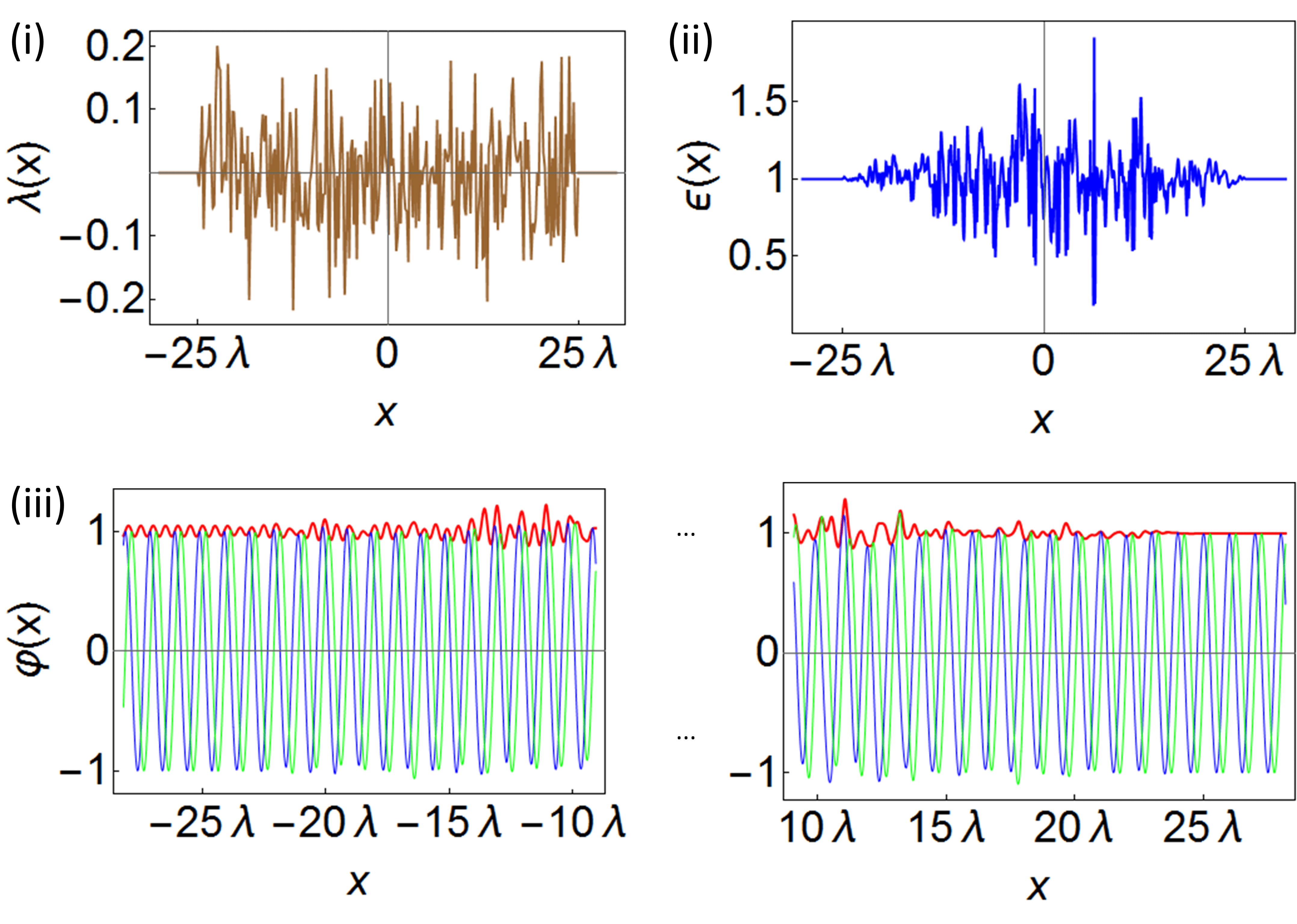}
        \caption{(i) The 'noise' \(\lambda(x)\) added onto the reflectionless permittivity profile. (ii) The new permittivity profile \(\epsilon(x)\) with the noise included. (iii) Time-averaged (red, upper) amplitude of a left incident wave of wavenumber \(\kappa\), propagating through a medium with permittivity given by (ii). Real and imaginary parts of the wave are shown in blue and green, respectively. The oscillations in the field norm to the left of the profile indicate a small amount of reflection.\label{noise}}
    \end{figure}

\textit{Sensitivity of perfect transmission to realising the exact design}
It has been discussed how the effect of slightly changing the wavenumber or the angle of incidence affects the transmission. However, it is not clear what the effect of slightly perturbing the permittivity itself. It is inevitable that there will be slight differences between the desired permittivity profile and that produced in a lab. To get an idea of the effect of this, consider adding some white noise to the permittivity of figure~\ref{BerryHowl} and seeing the behaviour of the wave through this new medium. This is shown in figure~\ref{noise}.
Unsurprisingly adding noise does have an effect on the transmissivity of the profile, since now the ansatz (\ref{9}) will no longer describe the correct solution (or, in the case of the spatial Kramers-Kronig media, the Kramers-Kronig relations will be violated). However, the more accurate the permittivity, the smaller the amplitude of the noise and therefore the smaller the reflection coefficient. i.e. The reflection and transmission coefficients will vary continuously with any continuous change to the permittivity profile. In order to make this precise, we define the amount of error in the profile due to the noise to be defined as
    \begin{equation}
        \left(\frac{\int dx\lambda^{2}(x)}{\int dx\chi^{2}(x)}\right)^{1/2}\label{S6},
    \end{equation}
where \(\lambda\) is the noise function and \(\chi\) is the susceptibility. We then calculated the reflection coefficients obtained from the wave propagation through ten different permittivity profiles generated in the same way as the example in figure~\ref{BerryHowl} and added ten different white noise functions in turn (each of the same standard deviation) to each and found that an average 7.47\(\%\) error led to an average reflection coefficient of 0.0159. Doubling the standard deviation and performing the same calculation resulted in an average 15.2\(\%\) error and this led to an average reflection coefficient of 0.0540. It is quite noticable that the reflection coefficient remains low for fairly significant errors in the permittivity and therefore is not just a sharp resonance type behaviour; rather the design method of the profile is fairly robust to deviations. A similar calculation was performed on the Kramers-Kronig profiles like that shown in figure~\ref{Fourier} and the sensitivity to deviations was similar. Namely an average 7.16\(\%\) error led to an average reflection coefficient of 0.0392 and an average 14.3\(\%\) error led to an average reflection coefficient of 0.0701.

\begin{figure}[ht!]
        \centering
	\includegraphics[width=\linewidth]{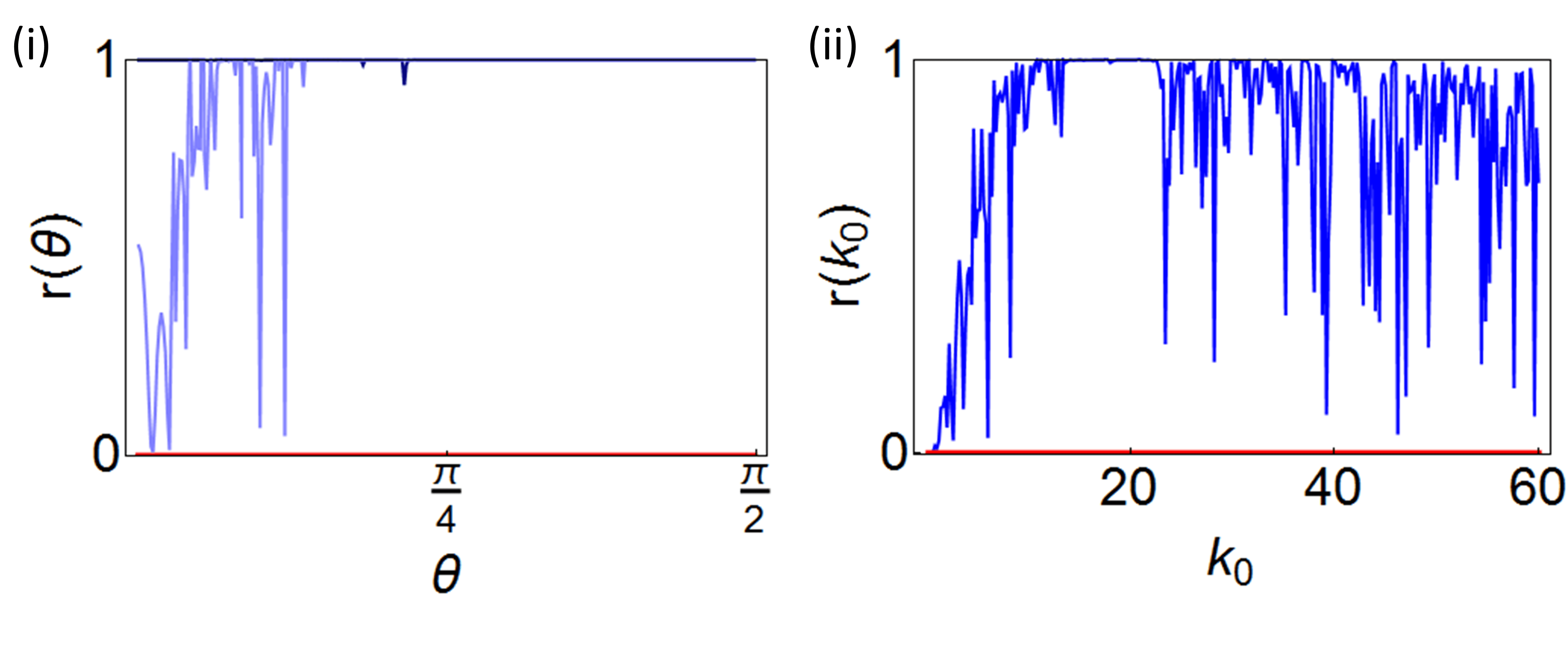}
        \caption{Demonstration of the importance of the Kramers-Kronig relations for suppressing reflection. (i) The reflection coefficient as a function of incidence angle for the Kramers-Kronig profile of figure~\ref{Fourier} at wavenumber \(5\pi\) (red) and for the profile consisting only of the corresponding real part at wavenumber \(5\pi\) (dark blue) and at wavenumber \(15\pi\) (light blue). (ii) The reflection coefficient as a function of wavenumber at normal incidence for the Kramers-Kronig medium (red) and its corresponding real part (blue).\label{reflection1}}
    \end{figure}

\begin{figure}[ht!]
        \centering
	\includegraphics[width=\linewidth]{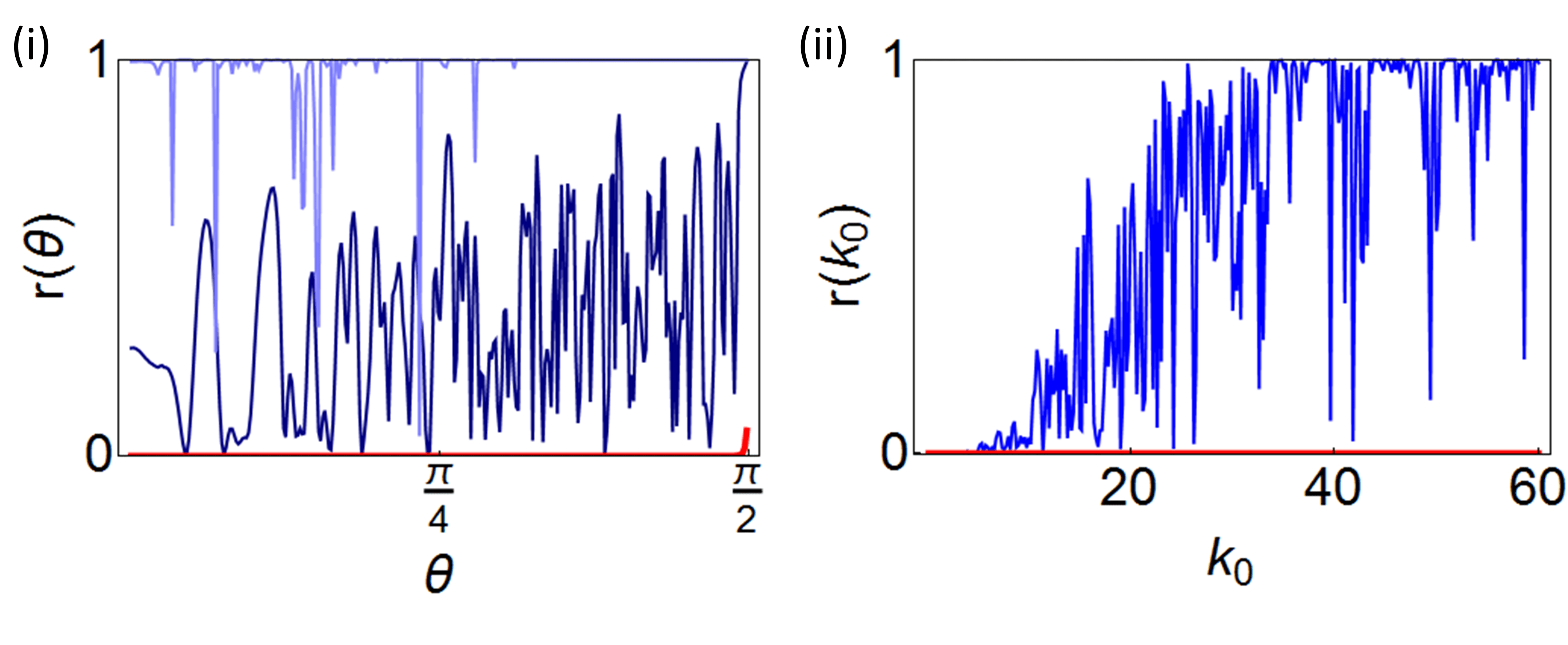}
        \caption{Demonstration of the importance of the Kramers-Kronig relations for suppressing reflection. (i) The reflection coefficient as a function of incidence angle for the Kramers-Kronig profile shown in figure~\ref{Pole} at wavenumber \(5\pi\) (red) and for the profile consisting only of the corresponding real part at wavenumber \(5\pi\) (dark blue) and at wavenumber \(15\pi\) (light blue). (ii) The reflection coefficient as a function of wavenumber at normal incidence for the Kramers-Kronig medium (red) and its corresponding real part (blue).\label{reflection2}}
    \end{figure}

\begin{figure}[ht!]
        \centering
	\includegraphics[width=\linewidth]{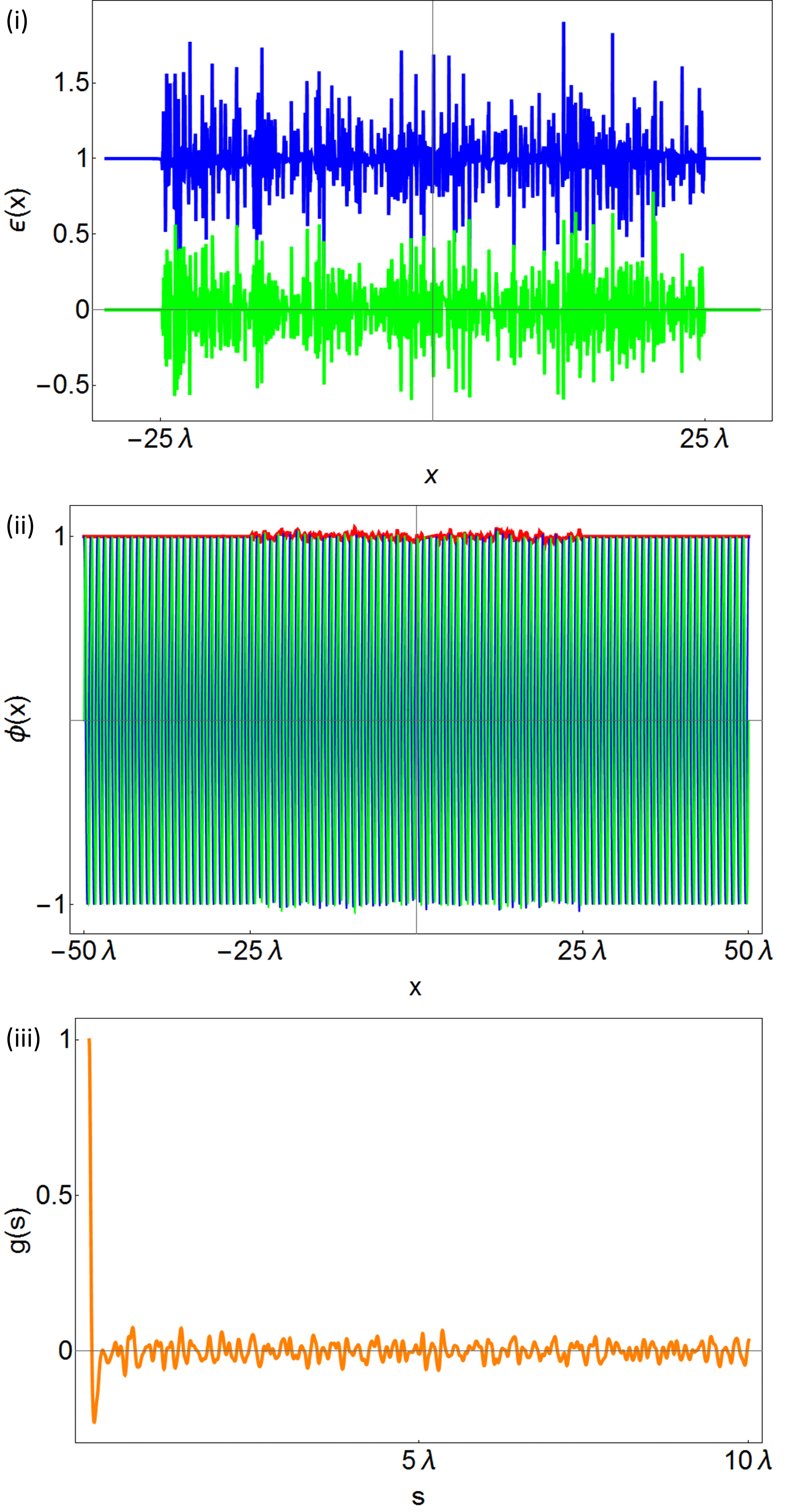}
        \caption{(i): A permittivity profile given by (\ref{S20}) with \(N=500\). (ii) The corresponding time-averaged amplitude (red) for a  left incident wave, \(\varphi\), with the real and imaginary parts of the wave are shown in blue and green, respectively. Again, the solution is seen to be reflectionless with perfect transmission. (iii) The correlation function of the permittivity, with correlation length \(x_{L}=0.0258\lambda\). Its Hurst exponents are \(H=0.152\) (real part) and \(0.182\) (imaginary part).\label{Pole}}
    \end{figure}

\textit{Wavenumber and Angle dependence}
Of the two recipes for generating reflectionless and perfectly transmitting media, the real-valued permittivity profiles only function at a single wavenumber and at normal incidence, whereas the complex-valued Kramers-Kronig permittivity profiles remain reflectionless at any wavenumber and at any incident angle (away from grazing incidence). This is confirmed numerically in figure~\ref{reflection1} for the Kramers-Kronig medium of figure~\ref{Fourier} discussed in the main part of the paper and in figure~\ref{reflection2} for the Kramers-Kronig medium of figure~\ref{Pole}.
These graphs show that, for general disordered media the reflection will be significant (often close to 100\%) for most frequencies and most angles of incidence whereas the spatial Kramers-Kronig media has negligible reflection over all frequencies and angles of incidence. This is despite having similar levels of disorder (indeed the correlation functions, and hence the correlation lengths, of the two media are identical). Including the imaginary part turns a general disordered medium, with the expected high reflection, into a medium whose susceptibility satisfies the spatial Kramers-Kronig relations, which are known to be reflectionless for all wavenumbers and all angles of incidence~\cite{Horsley2015,Horsley2016}.

\textit{Correlation function of a compact Fourier series:}
In the paper it is stated that the mean and variance of the correlation function of the finite compact Fourier series
    \begin{equation}
        \chi_{R}(x)=
    \begin{cases}
        \sum_{n=1}^{N}a_{n}\text{sin}\left(\frac{n\pi x}{L}\right),\qquad 0<x<L\\
        0,\qquad\qquad\qquad\qquad\qquad\text{otherwise}\label{S7}
    \end{cases}
    \end{equation}
vanishes in the limit \(N\to\infty\) for all positive \(s\), when the coefficients \(a_{n}\) are taken from a symmetric uniform distribution for \(n>1\) and \(a_{1}=-\sum_{m=1}^{\lfloor{\frac{N-1}{2}\rfloor}}\frac{a_{2m+1}}{2m+1}\). To show this, first write the mean as
    \begin{equation}
    \begin{split}
        \mu(g_{R}(s))&=\frac{2}{L}\sum_{n,m}g_{n,m}(s)\mu\left(\frac{a_{n}a_{m}}{\sum_{k}a_{k}^{2}}\right)\label{S8}
    \end{split}
    \end{equation}
where
    \begin{equation}
    g_{n,m}(s)=\left\{\begin{array}{ll}
         \frac{L\left(n\text{sin}\left(\frac{m\pi s}{L}\right)-(-1)^{m+n}m\text{sin}\left(\frac{n\pi s}{L}\right)\right)}{(n^{2}-m^{2})\pi^{2}},\qquad &n\neq m\\
         \frac{\left(n\pi(L-s)\text{cos}\left(\frac{n\pi s}{L}\right)+L\text{sin}\left(\frac{n\pi s}{L}\right)\right)}{2n\pi},\qquad\quad &n=m.\end{array}\right.\label{S9}
    \end{equation}
and
    \begin{equation}
    \begin{split}
        \mu\left(\frac{a_{n}a_{m}}{\sum_{k}a_{k}^{2}}\right)=0\qquad\text{for }n\neq m\label{S10}
    \end{split}
    \end{equation}
since the uniform random variable is even in \(a_{n}\) when \(n>1\). Also, by using
    \begin{equation}
    \begin{split}
        1&=\sum_{n=1}^{N}\mu\left(\frac{a_{n}^{2}}{\sum_{k=1}^{N}a_{k}^{2}}\right)\label{S11}
    \end{split}
    \end{equation}
 and substituting in \(a_{1}=-\sum_{m=1}^{\lfloor{\frac{N-1}{2}\rfloor}}\frac{a_{2m+1}}{2m+1}\), it follows that, at worst, \(\mu\left(\frac{a_{n}^{2}}{\sum_{k}a_{k}^{2}}\right)=\text{O}\left(\frac{1}{N}\right)\) as \(N\to\infty\). Meanwhile \(\sum_{n=1}^{N}g_{n,n}(s)\) remains bounded for \(s>0\) as \(N\to\infty\). Hence
    \begin{equation}
    \begin{split}
        \lim_{N\to\infty}\mu(g_{R}(s))&=0\qquad\text{for }s>0.\label{S12}
    \end{split}
    \end{equation}
    \par
The method used can be extended to show that the variance too vanishes in the same limit. Using the same notation, we can write the mean of the square of the correlation function as
    \begin{equation}
        \mu(g_{R}(s)^{2})=\frac{4}{L^{2}}\sum_{\alpha,\beta,\gamma,\delta=1}^{N}g_{\alpha,\beta}(s)g_{\gamma,\delta}(s)\mu\left(\frac{a_{\alpha}a_{\beta}a_{\gamma}a_{\delta}}{\left(\sum_{n=1}^{N}a_{n}^{2}\right)^{2}}\right)\label{S13}
    \end{equation}
Again the evenness of the uniform distribution can be used to eliminate all terms except those containing only squares of the coefficients
    \begin{equation}
    \begin{split}
        \frac{L^{2}}{4}\mu(g_{R}(s)^{2})&=\sum_{\alpha}g_{\alpha,\alpha}(s)^{2}\mu\left(\frac{a_{\alpha}^{4}}{\left(\sum_{n=1}^{N}a_{n}^{2}\right)^{2}}\right)\\
        &\quad+\sum_{\alpha\neq\beta}g_{\alpha,\alpha}(s)g_{\beta,\beta}(s)\mu\left(\frac{a_{\alpha}^{2}a_{\beta}^{2}}{\left(\sum_{n=1}^{N}a_{n}^{2}\right)^{2}}\right)\\
        &\quad+\sum_{\alpha\neq\beta}g_{\alpha,\beta}(s)^{2}\mu\left(\frac{a_{\alpha}^{2}a_{\beta}^{2}}{\left(\sum_{n=1}^{N}a_{n}^{2}\right)^{2}}\right)\label{S14}
    \end{split}
    \end{equation}
First consider the case when \(a_{1}\) is chosen from the same uniform distribution independently from the other coefficients. Expanding out the identity
    \begin{equation}
        1=\mu\left(\frac{\left(\sum_{n=1}^{N}a_{n}^{2}\right)^{2}}{\left(\sum_{n=1}^{N}a_{n}^{2}\right)^{2}}\right)\label{S15}
    \end{equation}
can be used to show that \(\mu\left(\frac{a_{\alpha}^{4}}{\left(\sum_{n=1}^{N}a_{n}^{2}\right)^{2}}\right)\) and \(\mu\left(\frac{a_{\alpha}^{2}a_{\beta}^{2}}{\left(\sum_{n=1}^{N}a_{n}^{2}\right)^{2}}\right)\) are O\(\left(\frac{1}{N^{2}}\right)\) as \(N\to\infty\). Meanwhile \(\sum_{\alpha}g_{\alpha,\alpha}(s)^{2}\), \(\sum_{\alpha\neq\beta}g_{\alpha,\alpha}(s)g_{\beta,\beta}(s)\) and \(\sum_{\alpha\neq\beta}g_{\alpha,\beta}(s)^{2}\) are each O\((N)\) as \(N\to\infty\). Combining these expressions shows that \(\mu(g_{R}(s)^{2})\to 0\) as \(N\to\infty\) for \(s>0\). The case when \(a_{1}=-\sum_{m=1}^{\lfloor{\frac{N-1}{2}\rfloor}}\frac{a_{2m+1}}{2m+1}\) is merely an additional technicality leading to more terms needing to be dealt with separately:
    \begin{equation}
    \begin{split}
        \frac{L^{2}}{4}\mu(g_{R}&(s)^{2})=\sum_{\alpha,\beta,\gamma,\delta=1}^{N}g_{\alpha,\beta}(s)g_{\gamma,\delta}(s)\mu\left(\frac{a_{\alpha}a_{\beta}a_{\gamma}a_{\delta}}{\left(\sum_{n=1}^{N}a_{n}^{2}\right)^{2}}\right)\\
        =&g_{1,1}(s)^{2}\mu\left(\frac{a_{1}^{4}}{\left(\sum_{n=1}^{N}a_{n}^{2}\right)^{2}}\right)+...\\
        &+\sum_{k,l,m,n=2}^{N}g_{k,l}(s)g_{m,n}(s)\mu\left(\frac{a_{k}a_{l}a_{m}a_{n}}{\left(\sum_{n=1}^{N}a_{n}^{2}\right)^{2}}\right)\label{S16}
    \end{split}
    \end{equation}
where the omitted terms correspond to some, but not all, of the coefficients \(\alpha,\beta,\gamma,\delta\) being 1. The last term of (\ref{S16}) is known to vanish in the \(N\to\infty\) limit for \(s>0\) owing to the previous analysis. Meanwhile
    \begin{equation}
    \begin{split}
        \mu&\left(\frac{a_{1}^{4}}{\left(\sum_{n=1}^{N}a_{n}^{2}\right)^{2}}\right)=\mu\left(\frac{\left(\sum_{m=1}^{\lfloor{\frac{N-1}{2}\rfloor}}\frac{a_{2m+1}}{2m+1}\right)^{4}}{\left(\sum_{n=1}^{N}a_{n}^{2}\right)^{2}}\right)\\
        &=\sum_{k,l,m,n=1}^{\lfloor{\frac{N-1}{2}\rfloor}}\frac{\mu\left(\frac{a_{k}a_{l}a_{m}a_{n}}{\left(\sum_{n=1}^{N}a_{n}^{2}\right)^{2}}\right)}{(2k+1)(2l+1)(2m+1)(2n+1)}\label{S17}
    \end{split}
    \end{equation}
However, we know that \(\mu\left(\frac{a_{k}a_{l}a_{m}a_{n}}{\left(\sum_{n=1}^{N}a_{n}^{2}\right)^{2}}\right)=\text{O}\left(\frac{1}{N^{2}}\right)\). On the other hand \(\sum_{k=1}^{N}\frac{1}{2k+1}\) increases as a \(log\) as \(N\to\infty\), so the expression in (\ref{S16}) will tend to 0. A similar argument can be used to show that the remaining terms also vanish in the large \(N\) limit and hence shows that
    \begin{equation}
        \lim_{N\to\infty}\sigma^{2}(g_{R}(s))=0\label{S18}
    \end{equation}
Therefore, although individual realisations of permittivity profiles generated using this method may have small, apparently random correlations, these will always go away as the number of terms is increased.

\begin{figure}[ht!]
        \begin{center}
	\includegraphics[width=\linewidth]{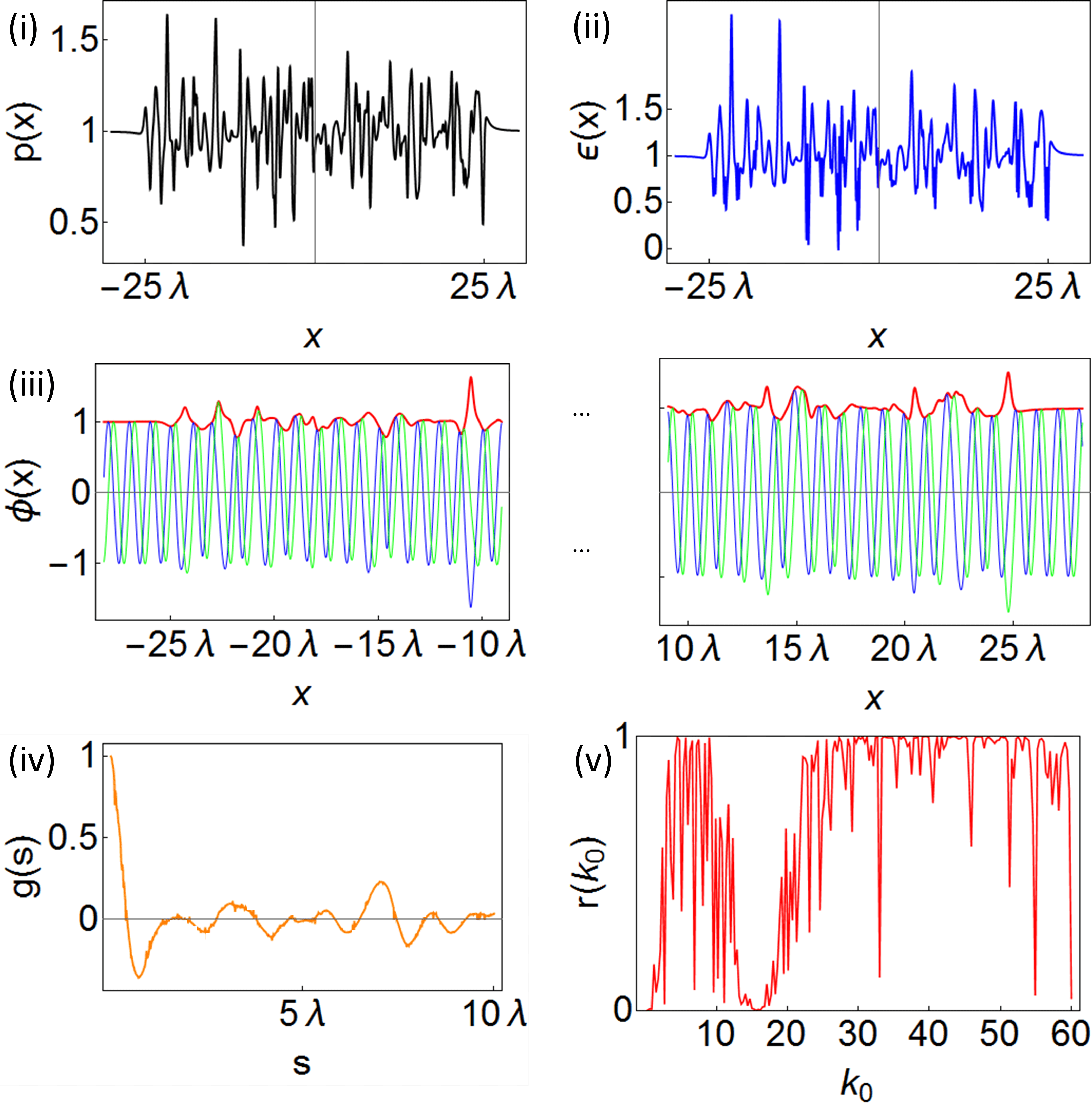}
        \caption{(i) A particular choice of \(p(x)\) given by (\ref{S19}) with \(N=250\). (ii) The corresponding permittivity when \(\kappa=5\pi\), having Hurst exponent \(H=0.224\). (iii) Time-averaged (red, upper) amplitude of a left incident wave of wavenumber \(\kappa\), propagating through a medium with permittivity \(\epsilon(x,\kappa)\). Real and imaginary parts of the wave are shown in blue and green, respectively. The lack of oscillations in (iii) indicates that the profile is reflectionless for this wave. The wave is also transmitted without a change in amplitude or a shift in phase. (iv) The profile's correlation function, with correlation length \(x_{L}=0.289\lambda\). (v) The reflection as a function of wavenumber \(k_{0}\). The reflection coefficient is very sensitive to the frequency of the incident wave and high reflection is possible except in a region about \(k_{0}=\kappa\) where the reflection is negligible.\label{PoleBerry}}
        \end{center}
    \end{figure}

\textit{Permittivity profiles constructed from poles:}
We have considered disordered  permittivity profiles constructed from finite Fourier series with randomly chosen coefficients. However, this is by no means the only way to generate disordered profiles. An alternative way was alluded to in the introduction: having poles of order two or higher in the lower half position plane. To generate real profiles using this method, take the function \(p(x)\) to be of the form
    \begin{equation}
        p(x)=1+\sum_{k=1}^{N}\frac{b_{k}}{(z-z_{k})^{2}}\label{S19}
    \end{equation}
where Im\((z_{k})<0\) and construct \(\epsilon\) using equation (\ref{10}) in the main part of the paper. A particular realisation is shown in figure~\ref{PoleBerry}.
Alternatively, one can construct corresponding complex-valued permittivity profiles by taking the real part of the susceptibility as a sum of double poles in the lower half plane
    \begin{equation}
        \chi_{R}(x)=1+\sum_{k=1}^{N}\frac{b_{k}}{(z-z_{k})^{2}}\label{S20}
    \end{equation}
and forming its imaginary part from the Hilbert transform of the real part, thus making a spatial Kramers-Kronig medium. An example is shown in figure~\ref{Pole}
Although there may appear little difference in the functions describing the permittivity profiles of a Fourier series with random coefficients and a series of double poles in the lower half plane, there is a difference in the local structure which is not obvious from their correlation functions.
    \par
The Hurst exponent of the sum of double poles is significantly lower than 0.5, indicating a long term negative correlation in the profile as seen in both figures. This is due to the local shape of such a function- a single double pole always has a region of high value following a region of low value in both the real and imaginary parts (and vice versa). However, for the Fourier series, this need not be the case due to the varying length scales on which the terms of the series oscillate, and thus there is little long term correlation in such a function, yielding a Hurst exponent closer to 0.5 (see figure~\ref{Fourier} in the main part of the paper).
    \par
The measures of disorder discussed in this paper, the correlation length and the Hurst exponent, depend continuously on the permittivity, and hence continuously on the design parameters (\(a_{n},b_{n},z_{n}\)). Due to the flexibility in the choice of these parameters, it is possible to obtain perfectly transmitting profiles with any desired correlation length and any Hurst exponent between 0 and 1. The correlation length can be tuned by changing the distribution that the parameters are randomly chosen from. Meanwhile the Hurst exponent can be tuned by changing the level of randomness in the design parameters- regular ordered structures, designed by choosing the parameters in a regular way, have a Hurst exponent which is very low (close to 0) or very high (close to 1). By then introducing some randomness in the choice of the parameters, the Hurst exponent shifts towards 0.5 i.e. more like a function generated from a random-walk like procedure.


\begin{thebibliography}{99}
\bibitem{footnote1} Note that this averaging process assumes the 'random phase model'--- not only should the transmission amplitudes be randomly distributed, but also their phases should be uniformly distributed between 0 and 2\(\pi\)~\cite{Berry1997,Stone1983}. However, there may be discrete frequencies when this random phase model fails to hold (see~\cite{Anderson1980} for a fuller discussion).
\end{thebibliography}
\end{document}